\documentclass{article}
\usepackage{spconf,amsmath,graphicx}
\usepackage{lipsum}
\usepackage{multirow}
\usepackage{booktabs}
\usepackage{adjustbox}
\usepackage{subcaption}
\usepackage{multirow, makecell}
\usepackage{dcolumn}
\usepackage{cite}
\newcolumntype{d}[1]{D{.}{.}{#1}}

\title{Dual-Attention Neural Transducers for Efficient Wake Word Spotting in Speech Recognition}
\name{%
\begin{tabular}{@{}c@{}}
Saumya Y. Sahai$^*$,
Jing Liu\sthanks{Equal Contribution.},
Thejaswi Muniyappa,
Kanthashree M. Sathyendra,\\
Anastasios Alexandridis, 
Grant P. Strimel,
Ross McGowan,
Ariya Rastrow,\\
Feng-Ju Chang,
Athanasios Mouchtaris,
Siegfried Kunzmann
\end{tabular}}

\address{Amazon Alexa AI, USA}
\begin{document}
\ninept
\maketitle

\begin{abstract}
We present dual-attention neural biasing, an architecture designed to boost Wake Words (WW) recognition and improve inference time latency on speech recognition tasks. This architecture enables a dynamic switch for its runtime compute paths by exploiting WW spotting to select which branch of its attention networks to execute for an input audio frame. With this approach, we effectively improve WW spotting accuracy while saving runtime compute cost as defined by floating point operations (FLOPs). Using an in-house de-identified dataset, we demonstrate that the proposed dual-attention network can reduce the compute cost by $90\%$ for WW audio frames, with only $1\%$ increase in the number of parameters. This architecture improves WW F1 score by $16\%$ relative and improves generic rare word error rate by $3\%$ relative compared to the baselines.
\end{abstract}


\section{Introduction}
\label{sec:intro}
End-to-end (E2E) ASR systems such as connectionist temporal classification (CTC) \cite{graves2006connectionist}, listen-attend-spell (LAS) \cite{chan2016listen}, recurrent neural network transducer (RNN-T) \cite{graves2012sequence}, transformer transducer \cite{vaswani2017attention,tian2019self,yeh2019transformer,zhang2020transformer, chang2021context}, and their variants ConvRNN-T \cite{radfar2022convrnn}, conformer \cite{gulati2020conformer,li2021better} have become increasingly popular due to their superior performance over hybrid HMM-DNN systems, making them promising architectures for deployment in commercial virtual voice assistants. While hybrid models optimize the acoustic model (AM), pronunciation model (PM) and language model (LM) independently, E2E systems jointly optimize them to output word sequences directly from an input sequence. These fully neural E2E approaches are strong candidates for low resource settings due to their simplicity and unified compression capabilities. However, one of the major limitations of E2E ASR systems is that they have difficulty in accurately recognizing words that are uncommon in the paired audio-text training data, such as custom WW which are specified by the customer to address a virtual assistant (e.g. Ziggy, Hey Shaq), contact names, proper nouns, and other rare named entities \cite{sainath2018no,gourav2021personalization}. To address this issue, previous works \cite{chang2021context,sathyendra2022contextual} have proposed attention-based neural biasing which apply a biasing adapter mechanism by scoring the similarity of encoded audio representations with personalized catalog embeddings. Attention-based neural biasing is a promising approach to boost personalized entity names; however, due to its compute complexity by application on the audio encodings frame-by-frame, the incurred runtime latency challenges scalable deployment of these attention-based biasing networks for on-device systems with hardware constraints (e.g. limited memory bandwidth and CPU constraints). 

To address compute limitations for on-device ASR, model compression is a commonly used methodology. In general, model compression techniques can be divided into two categories: architecture modification and weight interpretation. The former reduces complex architectures to simplified alternatives while the latter interprets weights with low-bit representations. Our work belongs to the architecture modification category. Also in this category are CIFG \cite{greff2016lstm} which simplifies the LSTM structure \cite{hochreiter1997long} by merging the input and forget gates which results in 25\% fewer parameters; simple recurrent unit \cite {lei2017simple} introduces more efficient recurrent cells for Edge ASR; low-rank factorization \cite{povey2018semi}, bifocal \cite{macoskey2021bifocal}, dynamic encoders \cite{dynamEncoder2021}, amortized networks \cite{xie2022compute, macoskey2021amortized}, linformer \cite{wang2020linformer}, performer \cite{choromanski2020rethinking} and time-reduction layers \cite{chan2016listen,soltau2016neural,he2019streaming} which are suggested to reduce runtime latency. In the second category, quantization \cite{hubara2017quantized, choi2019accurate,nguyen2020quantization}, sparsity \cite{zhu2017prune,niculae2017regularized,roy2021efficient} are dominant paradigms used to interpret weights with lower-bit integer or sparse representations.


Our work is inspired by the bifocal neural transducer \cite{macoskey2021bifocal}, that contains two audio encoder networks which are dynamically pivoted at run time. One major difference in our work is the compute cost amortized Multi-Head Attention (MHA) \cite{vaswani2017attention} biasing networks designed to simultaneously boost custom WW and personalized entities. In contrast to vanilla neural biasing \cite{chang2021context,sathyendra2022contextual} which does not differentiate sentence segments, the proposed dual-attention network biases towards only WW embeddings at the sentence-beginning, and proper name embeddings (e.g. contact names, device names) at post-WW segments.





\section{Related Work}
\label{sec:prior_work}

\begin{figure*}[tp]
\centering
\includegraphics[width=1.00\linewidth]{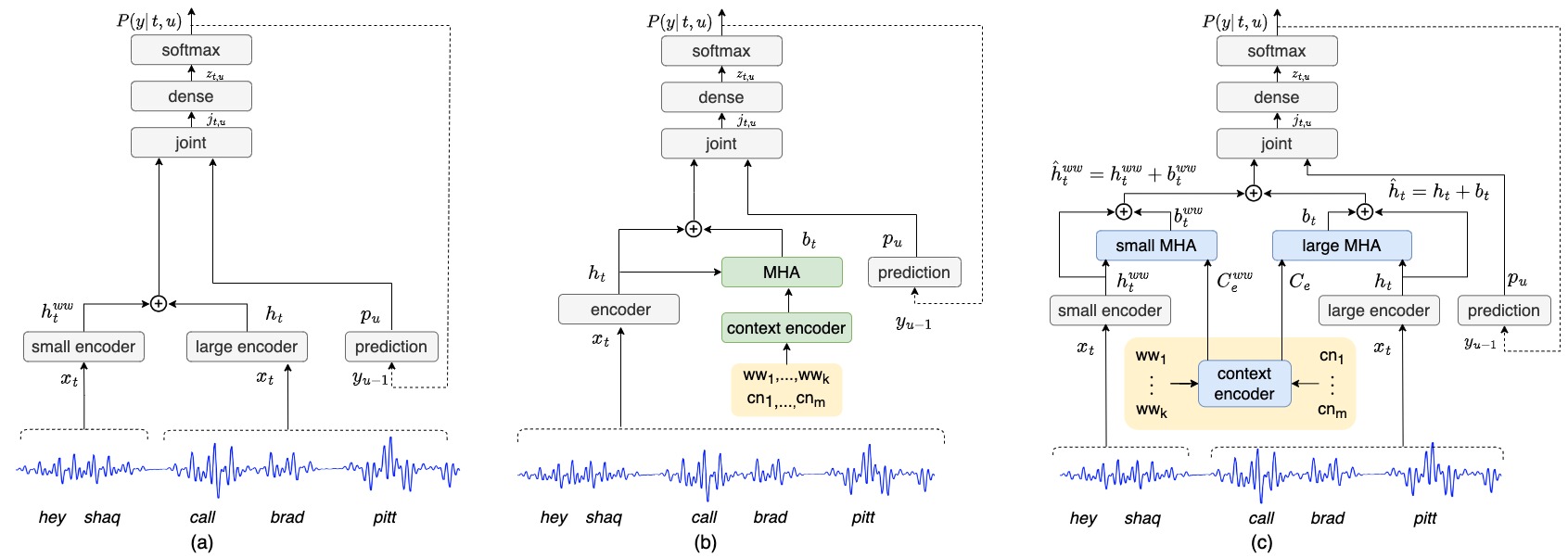}
\caption{Model architectures: (a) baseline bifocal neural transducer \emph{pretrained-base} \cite{macoskey2021bifocal}; (b) baseline single-attention neural biasing \emph{single-attn-base-128} \cite{chang2021context,sathyendra2022contextual}; (c) proposed dual-attention neural biasing \emph{dual-attn}-$\lambda$, where $\lambda$ is the projection size of the small MHA; initialize the model with pretrained weights (grey blocks); add dual-attention modules (blue blocks); only train the blue blocks by freezing the grey ones.}
\label{fig:dual_proposed}
\vspace{-2.5mm}
\end{figure*}

\subsection{Bifocal RNN-T}
\label{sec:bifocal}
Neural sequence transducers are streaming E2E ASR systems \cite{graves2012sequence} that typically consist of an audio encoder, a text predictor and a joint network. The encoder, behaving like an AM, produces high level acoustic representations $\mathbf{h}_t^{enc}$ for each input audio frame $\mathbf{x}_T=(x_0,\dots, x_T)$. The text predictor, acting like an LM, encodes previously predicted word-pieces $\mathbf{y}_{u-1}=(y_0,\dots,y_{u-1})$ and outputs $\mathbf{h}_u^{pred}$, with
\[
\mathbf{h}_t^{enc} = \text{AudioEncoder}(\mathbf{x}_t); \quad \mathbf{h}_u^{pred} = \text{TextPredictor}(\mathbf{y}_{u-1}).
\]
The joint network fuses $\mathbf{h}_t^{enc}$ and $\mathbf{h}_u^{pred}$ and passes them through dense and then softmax layers to obtain output probability distributions over the word-pieces. 

Bifocal RNN-T \cite{macoskey2021bifocal}, is a special type of neural transducer, consisting of two audio encoders: a small/fast encoder trained for the buffered lead-in audio segments that contains pre-WW and WW audio frames; and a large/slow encoder for processing the remainder of the audio leveraging WW spotting to pivot between the two (Fig.\ \ref{fig:dual_proposed}(a)). Bifocal architecture improves latency by diverting WW audio frames to its fast encoder branch. However, it has limited capacity to adapt itself to recognize new or rare words, particularly user-specified WW directed to its low-capacity small audio encoder. This drawback degrades user experience by falsely rejecting voice queries initiated from custom WW or diverting to another virtual assistant personality by mistake.
The proposed dual-attention architecture is designed to mitigate this limitation using an attention network to ``just focus on" boosting the desired WW.

\subsection{Neural Biasing}
\label{sec:nb}
To leverage a user's custom environment and preferences to improve recognition of personalized requests directed to voice assistants, both \cite{chang2021context} and \cite{sathyendra2022contextual} suggest neural biasing (Fig.\ \ref{fig:dual_proposed}(b)) consisting of MHA layers \cite{vaswani2017attention} to measure the similarity of audio encoding with entity-name embedding. The attention weights are computed frame-by-frame to assess the relevance of user pre-defined entity-names with the current audio frame. Neural biasing effectively boosts personalized entity-names (e.g. \emph{proper names} such as contacts and device names) because more relevant entity-names receive higher attention weights. However, due to quadratic complexity of dot-product attention \cite{vaswani2017attention}, runtime latency has been a bottleneck to deploy neural biasing to embedded ASR systems.

 \section{Dual-Attention Neural Biasing}
\label{sec:dual-attn}

\begin{table*}
\small
\centering
\resizebox{2\columnwidth}{!}{
\begin{tabular}{cccccccccc}
\toprule 
\multirow{2}{*}{Model} &\multirow{2}{*}{\makecell{Lead-in audio\\ segment FLOPs}}  & \multirow{2}{*}{\makecell{Biasing layer \\parameters}}& \multirow{2}{*}{\makecell{Lead-in audio\\catalog size}} &  \multicolumn{3}{c}{RNN-T}  & \multicolumn{3}{c}{C-T}  \\
\cmidrule(lr){5-7} 
\cmidrule(lr){8-10}
 &  & &  &  F1R & TRRR & TARR & F1R & TRRR &  TARR\\
\midrule
\emph{pretrained-base} & $-$ & $-$ & $-$ & $-$ & $-$ & $-$ & $-$ & $-$ & $-$\\
 \emph{single-attn-base-128}  & $3.3M $  & $400K$\footnotesize{ } & 300 & +14.72\%  & -14.80\% & +28.73  \% & +14.96\%  & -12.84\% & +41.96\% \\
 \emph{single-attn-catalog-mask} & $190K$  \footnotesize{(-93.7\%)} & $400K$\footnotesize{(+0.0\%)} & 6 & +17.03\% & -15.17\% & +31.39\% & +16.93\% & -11.11\% & +42.77\% \\
 \emph{dual-attn-64}      & $95K$ \footnotesize{(-96.9\%)}&  $483K$\footnotesize{(+20.7\%)} & 6 & +16.42\% & -14.07\% &   +36.45\% & +14.14\%  & -7.98\% & +32.42\% \\
 \emph{dual-attn-32}      & $48K$  \footnotesize{(-98.4\%)}&  $441K$\footnotesize{(+10.3\%)} & 6 & +16.00\% & -12.43\% &   +26.07\% & +15.34\%  & -12.15\% & +42.51\% \\
 \emph{dual-attn-16}     & $23K$  \footnotesize{(-99.2\%)}&  $421K$\footnotesize{(+5.1\%)}& 6 & +14.34\% & -16.63\% &  +29.01\% & +14.38\% & -12.15\% & +39.23\% \\
 \emph{dual-attn-8}       & $12K$  \footnotesize{(-99.6\%)}&  $410K$ \footnotesize{(+2.5\%)}& 6 & +2.38\% & -4.21\% &  +8.52\% & +9.30\% & -5.90\% & +19.07\% \\
 \bottomrule
\end{tabular}}
\caption{Compute cost measured in FLOPs ($M$=$10^6$, $K$=$10^3$); number of parameters in the biasing layers;  relative changes in F1 score (denoted as F1R), True Reject Rate (TRRR) and True Accept Rate (TARR) for the proposed \emph{dual-attn}-$\lambda$ in Fig.\ \ref{fig:dual_proposed}(c), comparing with baseline \emph{pretrained-base} as shown in Fig.\ \ref{fig:dual_proposed}(a); \emph{single-attn-base-128} refers to Fig.\ \ref{fig:dual_proposed}(b); +/- sign means to improve/degrade.}
\label{tab:ww-metrics-all}
\end{table*}

To address the on-device latency and compute limitations, and inspired by bifocal RNN-T and neural biasing (Sec.\ \ref{sec:prior_work}), we propose dual-attention neural biasing (Fig.\ \ref{fig:dual_proposed}(c)), which enables a dynamic pivot for its runtime compute paths, namely leveraging WW spotting to select the branch of the network to execute an input audio frame on.
The motivation of dual attention is to introduce bifocal ``lenses" engineered to focus on different segments of an utterance. The distinguishing feature of this design is training two alternative MHA networks (highlighted blue components in Fig.\ \ref{fig:dual_proposed}(c)). A small attention network $\mathcal{A}^s$, coupled with the small audio encoder, is trained for the lead-in segments and a large attention network $\mathcal{A}^l$ paired with the large audio encoder for the rest of the audio. $\mathcal{A}^s$ is designed to boost the ASR accuracy for user-specified custom WW (typically in the order of 10), while $\mathcal{A}^l$ is engineered to improve the recognition of personalized entity names (can scale to tens of thousands). $\mathcal{A}^s$ has a smaller number of hidden units than $\mathcal{A}^l$, enabling faster but coarser frame processing. In contrast, $\mathcal{A}^l$ has a larger capacity, but at the cost of more compute. The final component in this design is the context encoder, namely a BiLSTM encoder, which takes tokenized custom WW/proper names from a sentence-piece tokenizer \cite{kudo2018subword}. The last state of this BiLSTM is used as the embedding $C_e^{ww}$ or $C_e$. In Fig. \ref{fig:dual_proposed} (c), we first pretrain the bifocal transducer (grey blocks), then fine tune only the context encoder and the two MHA models (blue blocks) by keeping the rest of the pretrained weights (grey blocks) frozen \cite{sathyendra2022contextual}. 

\subsection{Dual-Attention Biasing Networks}
The small MHA network $\mathcal{A}^s$ is trained to learn the correlation between the lead-in audio encoding and user enabled WW text embedding. It is a light-weight model thanks to the natural lower perplexity of the spoken words prior to the WW. In contrast, higher perplexity of the post WW segment requires an MHA model $\mathcal{A}^l$ with higher capacity. The objective of this dual-attention design is to match the accuracy of single-attention baseline (Sec.\ \ref{sec:nb}) and to reduce the FLOPs since this architecture emphasizes the reduction of the MHA inference cost as it is one of the primary runtime bottlenecks. 





\subsection{Dynamic Catalog Masking}
\label{sec:catalog_masking}
In contrast to \cite{chang2021context,sathyendra2022contextual} (Fig.\ \ref{fig:dual_proposed}(b)) which statically concatenates WW and \emph{proper names} embeddings without differentiating sentence segments, one distinguishing feature in our design (Fig. \ref{fig:dual_proposed}(c)) is catalog masking which is dynamically determined by the
frame index signaling the end of the WW. At inference time, we only apply WW embeddings $C_e^{ww}$ in $\mathcal{A}^s$ by dynamically masking out \emph{proper names} tokens. In this way, we effectively narrow down the biasing candidates from tens of thousands (i.e. \emph{proper names}) to 6 (i.e. custom WW only) for better focusing, and to rule out less relevant catalogs like contacts/device names from appearing at the sentence-beginning. Similarly, for $\mathcal{A}^l$, we drop out WW catalogs and only apply \emph{proper names} embeddings $C_e$. In this way, we enable the two MHAs to bias toward their own targets by masking out irrelevant catalogs. As shown in Fig.\ \ref{fig:dual_proposed}(c), the highlighted yellow box containing the context encoder runs offline to generate and cache neural embeddings $C_e^{ww}$ for user-customized WW $ww_1, \dots, ww_k$ and $C_e$ for proper names $cn_1,\dots,cn_m$. These cached embeddings $C_e^{ww}$ and $C_e$ are dynamically masked for runtime inference. We also introduce a special \emph{no-bias} token into our catalog as in \cite{pundak2018deep,sathyendra2022contextual} to help the dual-attention system to learn when not to bias. 




\section{Experiments}
\label{sec:experiments}

\subsection{Datasets}
We use 114K hours of de-identified in-house voice assistant (\emph{general}) dataset randomly sampled from live traffic across more than 20 domains (e.g. Music, Communications, SmartHome) to pretrain the baseline RNN-T and Conformer-Transducer (C-T) models\footnote{As far as we know, there does not exist a large-scale public dataset that contains a variety of user-customized WW.}. For training the dual-attention networks, we use 290 hours of \emph{proper names} (that contains mentions of named entities), \emph{general} data which is mixed in the ratio of 1:2.5, and 3.6K hours of semi-supervised dataset containing 6 custom WW generated using a teacher model \cite{liu2021teacher}. To evaluate the models, we use a 75-hour \emph{general} testset and a 20-hour \emph{proper names} testset which are both human-transcribed. For calculating WW true accept and true reject rates, we use 25 hours of human annotated data containing 6 WW. The training and test sets are de-identified and have no overlap.

\subsection{Experimental Setup}
We evaluate the dual-attention neural transducers with two pretrained ASR architectures, RNN-T and C-T. 

\textbf{RNN-T and C-T Pretraining.} The input audio features are 64-dimensional LFBE features extracted every 10ms with a window size of 25ms resulting in 192 feature dimensions per frame. Ground truth tokens are passed through a 2.5K and 4K word-piece tokenizer \cite{sennrich2015neural,kudo2018subword} for RNN-T and C-T, respectively. The RNN-T encoder has 5 LSTM layers and a time reduction layer with downsampling factor of 2 at layer 3. Each LSTM layer has 256 units each layer for the lead-in audio encoder and 1120 units for the large audio encoder (Fig.\ \ref{fig:dual_proposed}(a)).
The prediction network has 2 LSTM layers with 1088 units each. The C-T encoder network consists of 2 convolutional layers with 128 kernels of size 3, and strides 2 and 1 for the first and second layer, respectively, followed by a dense layer to project input features to 512 dimensions. They are then fed into 14 conformer blocks, that contain layer normalizations and residual links between layers. Each conformer block has a 1024 unit feed-forward layer, 1 transformer layer with 8 64-dimensional attention heads and 1 convolutional module with kernel size 15. The prediction network has 2 LSTM layers with 736 units each. Convolutions and attentions are computed on the current and previous audio frames to make it streamable. For both RNN-T and C-T, the encoder and prediction network outputs are projected through 512 units of a feedforward layer.

\textbf{Baselines.} The baseline \emph{pretrained-base} shown in Fig.\ \ref{fig:dual_proposed}(a) is a bifocal neural transducer \cite{macoskey2021bifocal} (RNN-T or C-T)  which has two audio encoders but no neural biasing layers \cite{chang2021context,sathyendra2022contextual}. This model has fast inference but poor accuracy on \emph{proper names}. The second baseline \emph{single-attn-base-128}, displayed in Fig.\ \ref{fig:dual_proposed}(b), is a single-attention neural biasing transducer model as in \cite{chang2021context,sathyendra2022contextual} with keys and values projected to 128 dimensions. This baseline has good accuracy on \emph{proper names} but slower inference at runtime compared to the \emph{pretrained-base}. The third model \emph{single-attn-catalog-mask} is the same as \emph{single-attn-base-128} (Fig.\ \ref{fig:dual_proposed}(b)) except that irrelevant catalogs (e.g. \emph{proper names}) are removed from the small audio encoder via dynamic catalog masking (Sec. \ref{sec:catalog_masking}).



\textbf{Configuration for dual-attention biasing networks.} The context encoder is a BiLSTM with 64 units (for each forward and backward LSTM). The input and output have 64-dimensional projections. This context encoder is trained from scratch to generate embeddings for both WW and proper names. These embeddings are then fed into the dual-attention layers to bias the audio encoders outputs (Fig.\ \ref{fig:dual_proposed}(c)). More precisely, the large MHA network takes the large audio encoder outputs as query, and proper names embeddings as key and value and projects them to 128 dimensions. On the other hand, the small MHA is a light-weight model which takes query from the small audio encoder outputs, and WW embeddings as key and value and projects them to size $\lambda$, denoted as \emph{dual-attn}-$\lambda$. We experiment with $\lambda=64, 32, 16$, and $8$. In the following experiments, both small and large MHA only use 1 attention head, since we did not observe accuracy gains with 2 or 4 heads.

\section{results}
\label{sec:results}

Given a model A's WER (WER$_A$) and a baseline B’s WER (WER$_B$), the relative Word Error Rate Reduction (WERR) of A over B is computed as $\text{WERR}=(\text{WER}_B-\text{WER}_A)/\text{WER}_B$. The WW accuracy is measured in terms of True Accept Rate (TAR), True Reject Rate (TRR) and F1 score, where TAR is the proportion of ground truth positives that are accepted correctly; TRR is the fraction of ground truth negatives that are rejected correctly and F1 score is the harmonic mean of TAR and TRR. We present F1, TAR and TRR relative improvement (denoted by F1R, TARR and TRRR respectively) against a baseline model. Higher values of WERR, F1R, TARR and TRRR represent better performance. Negative values mean degradation. To measure compute cost, we report the total number of MHA-layer floating point operations \cite{macoskey2021bifocal,xie2022compute} per frame (FLOPs) required for the lead-in audio.

\vspace{-3mm}
\subsection{Wake word accuracy} As the model learns to bias towards the desired WW, TARs naturally improve. TRR may degrade since the model is biased to accept more queries. From Table \ref{tab:ww-metrics-all}, the proposed model \emph{dual-attention}-$\lambda$ improves WW TARR by up to $36\%$ (RNN-T) and $42\%$ (C-T) against their bifocal baselines \emph{pretrained-base}. As we reduce the projection dimension $\lambda$ from $64$ to $16$, we still see $29\%$ improvement of TARR for RNN-T, and $39\%$ for C-T. When $\lambda$=$8$, the TARR improvement is $8.5\%$ and $19\%$ for RNN-T and C-T respectively. On the other hand, we observed up to $14\%$ and $12\%$ regression in TRRRs for RNN-T and C-T. In fact, the proposed dual-attention model \emph{dual-attn}-$\lambda$ slightly outperforms Fig.\ \ref{fig:dual_proposed}(b) baseline \emph{single-attn-base-128} in TRRR for $\lambda=64, 32, 8$ for RNN-T and all values of $\lambda$ for C-T. Taking into account of both TAR and TRR, \emph{dual-attention}-$\lambda$ improves F1-score by up to $17\%$ for both RNN-T and C-T.


\vspace{-3mm}
\subsection{ASR accuracy} Table \ref{tab:WERR} presents the WERR for the proposed \emph{dual-attention}-$\lambda$ (Fig.\ \ref{fig:dual_proposed}(c)) against the bifocal baseline \emph{pretrained-base} (Fig.\ \ref{fig:dual_proposed}(a)). On \emph{proper names} test set, dual-attention architectures reduce WER by up to $28.6\%$ for RNN-T (vs. $26.7\%$ for \emph{single-attn-base-128}), and $30.4\%$ for C-T (vs. $29.1\%$ for \emph{single-attn-base-128}) thanks to catalog masking (detailed in Sec.\ \ref{sec:catalog_masking}) and a specialized MHA trained to ``focus on" more relevant catalogs. As we reduce $\lambda$ from 64 to 8 for the WW MHA network, we do not observe accuracy degradation against \emph{single-attn-base-128} on \emph{proper names} or \emph{general} test sets. This shows that \emph{dual-attn}-$\lambda$ reduces compute cost without hurting ASR accuracy for both RNN-T and C-T. 




\begin{table}
\centering
\resizebox{\columnwidth}{!}
{
\begin{tabular}{ccccc}
\toprule
& \multicolumn{2}{c}{RNN-T}  & \multicolumn{2}{c}{C-T}  \\
\cmidrule(lr){2-3}
\cmidrule(lr){4-5}
& \emph{general} & \emph{proper names} & \emph{general} & \emph{proper names} \\
\midrule
\emph{pretrained-base}          & $-$ &$-$ & $-$ & $-$ \\
 \emph{single-attn-base-128}    & +0.2\% & +26.73\% & +0.2\% & +29.08\% \\
\emph{single-attn-catalog-mask} & +0.2\% & +28.13\% & -0.2\% & +29.63\% \\
 \emph{dual-attn-64}            & +0.2\% & +28.13\% & 0.0\% & +29.23\% \\
 \emph{dual-attn-32}            & +0.8\% & +27.24\% & 0.0\% & +29.08\% \\
 \emph{dual-attn-16}            & +0.8\% & +27.98\% & 0.0\% & +30.42\% \\
 \emph{dual-attn-8}             & +0.2\% & +28.64\% &-0.2\% & +29.23\% \\
 \bottomrule
\end{tabular}}
\caption{WERR relative to \emph{pretrained{\text-}base}; +/- sign implies an improvement/degradation in WER.}
\label{tab:WERR}
\end{table}


\begin{figure}[tp]
\centering
  \centerline{\includegraphics[width=\columnwidth]{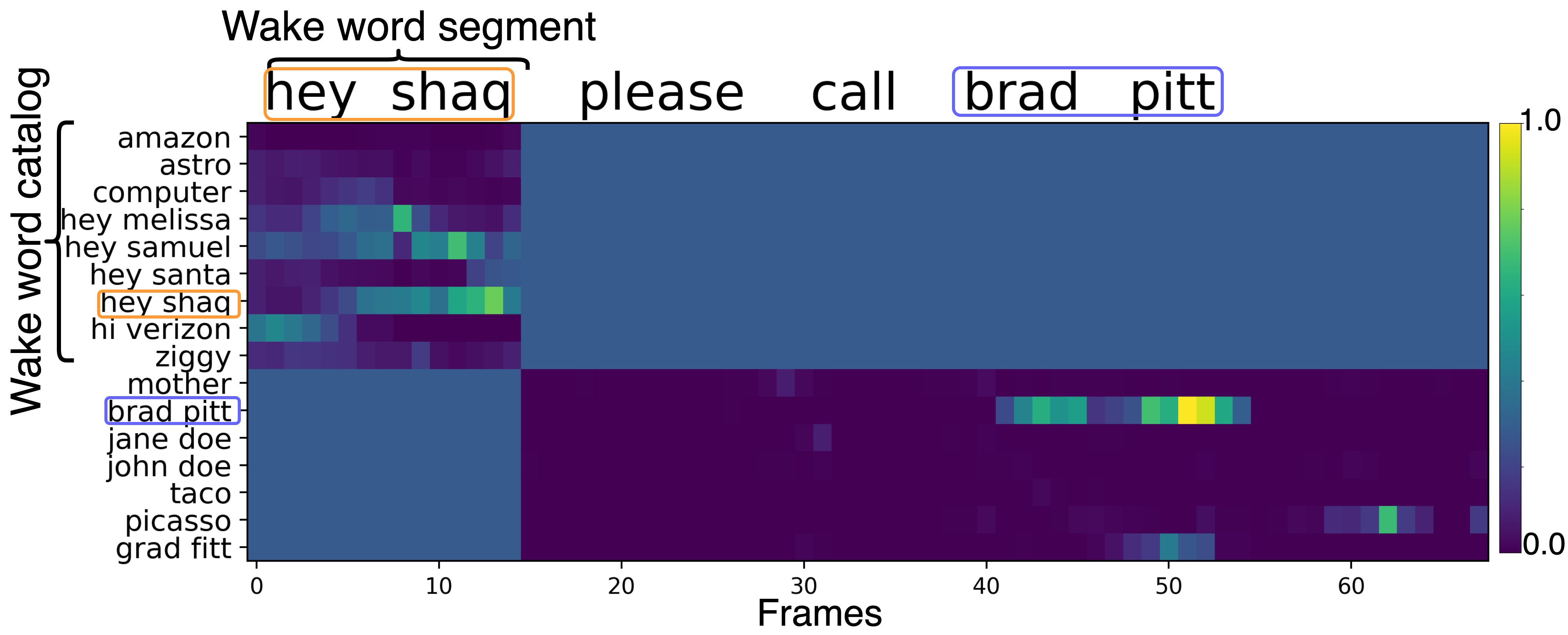}}
\caption{Attention weight visualization over different catalog entities for \emph{dual-attn-64} model (bright colors represent a greater weight). Each frame represents $60$ms.}
\label{fig:ww-attn}
\end{figure}

\vspace{-3mm}
\subsection{Compute Cost \& Attention Visualization} Table \ref{tab:ww-metrics-all} shows the compute cost measured in FLOPs. Using neural biasing \cite{chang2021context, sathyendra2022contextual} \emph{single-attn-base-128} as baseline (Fig.\ \ref{fig:dual_proposed}(b)), with a small increase of $83K$ parameters (from $400K$ to $483K$ or $1\%$ of ${\sim} 80M$ parameters), FLOPs (with catalog size 300) decreases from $3.3M$ (\emph{single-attn-base-128}) to $95K$ (\emph{dual-attn-64}). As we reduce $\lambda$ from 64 to 8, we further improve FLOPs from 95K to 12K. However, when $\lambda=8$, we observe reduced gain in WW accuracy: a relative F1 score improvement of 1\% and 6\% respectively for RNN-T and C-T. It is worth noting that catalog masking plays an important role in reduce FLOPs (from $3.3M$ to $190K$) as we narrow down the biasing candidates from 300 to 6 for the lead-in audio segment. In figure \ref{fig:ww-attn}, we visualize attention heat map for the small and large MHA layers with catalog masking. The small MHA layer shows high values for true WW \emph{hey shaq} while less relevant \emph{proper names} catalogs are masked out, whereas the large MHA biases towards the target proper name \emph{brad pitt} while the WW catalogs are masked out. 

\section{Conclusion}
\label{sec:conclusion}

We proposed a dual-attention neural transducer network which was inspired by bifocal RNN-T as well as attention-based neural biasing. This proposed architecture exploited WW spotting to dynamically select a biasing branch and efficiently boosted the ASR accuracy of proper names as well as custom WW, at the same time reducing runtime compute FLOPs and alleviated runtime latency of attention networks.

\vspace{-1mm}
\bibliographystyle{IEEEbib}
\bibliography{strings,refs}

\end{document}